\documentclass[letter,oneside,9pt,twocolumn,article]{aiaa}
\usepackage{graphicx}
\usepackage{rotating}
\usepackage{amssymb}
\usepackage{bm}
\usepackage{flushend}
\usepackage{cite}
\usepackage[super,comma]{natbib}
\usepackage[bf,normalsize]{subfigure}
\usepackage{adjustbox}
\usepackage{tabularx}
\usepackage{multirow}
\usepackage{multicol}
\usepackage{indentfirst} 
\usepackage[hidelinks]{hyperref}
\usepackage{float}
\usepackage{xcolor}
\usepackage{mwe}
\usepackage[markcase=noupper]{scrlayer-scrpage}
\usepackage{caption} 

\ihead*{{\it Physics of Fluids 38, 011705 (2026), \href{https://doi.org/10.1063/5.0314083}{10.1063/5.0314083}}}

\ohead*{\pagemark}
\ifoot*{}% left side of footer
\ofoot*{\rm Unrestricted content}% right side of footer
\author{
Bernard Parent\thanks{Associate Professor, author to whom correspondence should be addressed, bparent@arizona.edu.} 
~~and~ Felipe Martin Rodriguez Fuentes\thanks{Graduate Student}
\\[0.3em] \it University of Arizona, Tucson, AZ 85721, USA.
}

\title{
Thermodynamically Consistent Vibrational-Electron Heating: Generalized Model for Multi-Quantum Transitions
}

\abstract{ 
Accurate prediction of electron temperature ($T_{\rm e}$) is critical for non-equilibrium plasma applications ranging from hypersonic flight to plasma-assisted combustion. We recently proposed a thermodynamically consistent model for vibrational-electron heating [Phys. Fluids 37, 096141 (2025)] that enforces the convergence of $T_{\rm e}$ to the vibrational temperature ($T_{\rm v}$) at equilibrium. However, the original derivation was restricted to single-quantum transitions, limiting its validity to low-temperature regimes ($T_{\rm e} \lesssim 1.5$ eV). In this Letter, we generalize the model to include multi-quantum overtone transitions, extending its applicability to high-energy regimes. We demonstrate that previous models neglecting hot-band transitions incur a systematic heating error of $\exp(-\theta_{\rm v}/T_{\rm v})$, where $\theta_{\rm v}$ is the characteristic vibrational temperature. This error exceeds 40\% when $T_{\rm v}$ is greater than $\theta_{\rm v}$, effectively preventing thermal relaxation. To correct this, we derive a formulation where the total heating rate is a summation of channel-specific cooling rates $Q_{\rm e-v}^{(m)}$, each associated with a quantum jump $m$, scaled by a thermodynamic factor $\exp(m\theta_{\rm v}/T_{\rm e}-m\theta_{\rm v}/T_{\rm v})$. This generalized model preserves thermodynamic consistency by ensuring zero net energy transfer at equilibrium.
~\\
}

\nomenclature{
\begin{nomenclaturelist}{Roman symbols}

\item[$C_k$]{charge of $k$th  species, C}
  \end{nomenclaturelist}

  \begin{nomenclaturelist}{Greek symbols}
\item[$\beta_{k}^{\rm n}$]{parameter equal to 1 when $k$th species is neutral, 0 otherwise}
  \end{nomenclaturelist}

}
\begin{document}
\maketitle
%\makenomenclature

\dropword The~energy~exchange between free electrons and molecular vibrational modes, known as electron-vibrational (e-V) coupling, is a cornerstone process in numerous applications involving non-equilibrium plasmas. Its accurate representation is vital for modeling hypersonic flight, plasma-assisted combustion (PAC), and laser-induced plasmas (LIP).

In hypersonic flight beyond Mach 10, the plasma layer enveloping a vehicle critically influences its operation. This plasma enables advanced technologies such as electromagnetic shielding, as explored in early work by Musal\cite{gmrl:1963:musal} and later by Gregoire, \cite{dtic:1992:gregorie} as well as plasma antennas, \cite{ieee:2024:magarotto} electron transpiration cooling, \cite{aip:2017:hanquist,aiaapaper:2021:parent} and magnetohydrodynamic (MHD) systems. \cite{aiaaconf:2022:moses,jtht:2023:parent,aiaa:2025:parent} The viability of these technologies hinges on precise predictions of plasma properties like electrical conductivity and plasma frequency, which are strongly dependent on the electron temperature ($T_{\rm e}$). Indeed, the collision cross sections needed to determine electrical conductivity, for instance, are a function of electron temperature. Furthermore, electrical conductivity depends on the plasma density, which is also a function of $T_{\rm e}$ through processes like electron-ion recombination and ambipolar diffusion to surfaces. \cite{pf:2022:parent} Plasma density, in turn, governs the
plasma frequency, which is crucial for electromagnetic interactions. In the characteristic non-equilibrium of re-entry flows, where nitrogen is not fully dissociated, e-V coupling is the dominant mechanism governing $T_{\rm e}$. Specifically, superelastic collisions with vibrationally excited N$_2$ molecules become the primary heating source for electrons, yet direct data for these heating rates are scarce.

Similarly, e-V coupling is central to PAC. High-energy electrons create radicals and excited species while also initiating thermal pathways. These pathways include fast heating from the quenching of electronic states \cite{apl:2011:bak} and slow heating from the vibrational-translational (V-T) relaxation of N$_2$ molecules excited via e-V transfer. Detailed kinetic mechanisms \cite{rsta:2015:adamovich,jap:2023:chen} model this process, whereas phenomenological approaches often assume a fixed energy partition. \cite{cf:2016:castela} However, recent studies have shown this energy partition is not fixed and varies with operating conditions, \cite{cf:2025:dijoud} necessitating a predictive model for e-V energy transfer to accurately simulate PAC.

Laser-induced plasmas also require accurate e-V coupling models. \cite{jap:2019:peters, jap:2010:shneider, jap:2023:pokharel} In these systems, while electron-impact ionization and inverse Bremsstrahlung dominate energy exchange during the breakdown phase,\cite{book:1989:radziemski,jpd:2019:alberti,jcp:2020:munafo} electron-vibrational coupling becomes the dominant mechanism for both energy loss and gain in the inter-pulse phase where $T_{\rm e}$ drops below 3--5 eV. Furthermore, the reverse process---heating via superelastic collisions---plays a critical role throughout, particularly in multi-pulse regimes where accumulated vibrational energy slows the decay of $T_{\rm e}$. This is paramount because $T_{\rm e}$ controls the rates of electron attachment and recombination, which dictate the plasma's lifetime.

Modeling the vibrational-electron heating rate has been a persistent challenge. Early phenomenological models scaled the heating rate with simple temperature or energy ratios. \cite{jap:2010:shneider,jtht:2012:kim,jtht:2013:farbar} Although these drive the system towards equilibrium ($T_{\rm e} = T_{\rm v}$), they lack a rigorous foundation. More recent models applied the principle of detailed balance, \cite{jap:2019:peters,jap:2023:pokharel} but their formulation contained a crucial flaw that prevents the convergence of electron and vibrational temperatures at equilibrium, violating thermodynamic consistency.\cite{pf:2025:rodriguez:2}

To address these shortcomings, we recently developed a novel, thermodynamically consistent model for vibrational-electron heating in Ref.~\citenum{pf:2025:rodriguez:2}. That work established a robust heating-to-cooling ratio derived rigorously from detailed balance by assuming a Boltzmann distribution for vibrational states. However, the original derivation relied on the simplifying assumption that electron energy loss was dominated by fundamental (i.e.~single quantum)  transitions  from the ground state, thereby neglecting overtone processes. This constraint effectively restricted the model's validity to regimes where the electron temperature remains below approximately 1.5 eV, where single-quantum transitions dominate. In this Letter, we extend that proof to include electron cooling by vibrationally excited states and multi-quantum overtone transitions. This new proof extends the applicability of the referenced model to higher-energy regimes.

We derive this generalization from basic principles, ensuring the model satisfies the detailed balance principle and remains thermodynamically consistent. 
To extend the thermodynamically consistent model presented in the original work to regimes where overtone transitions are non-negligible, we must relax the assumption of a single effective activation energy and instead treat the fundamental and overtone processes as parallel channels.

While the original derivation aggregates all transitions into a single cooling flux, the generalized approach decomposes the macroscopic electron cooling rate $Q_{\rm e-v}$ into components based on the change in vibrational quantum number $m$. The total cooling is defined as
\begin{equation}
    Q_{\rm e-v} = \sum_{m=1}^\infty Q_{\rm e-v}^{(m)}
    \label{eq:total_cooling}
\end{equation}
where $Q_{\rm e-v}^{(m)}$ represents the electron energy loss due to a transition channel with a quantum number jump of $m$ and an energy gap $\Delta {\mathcal E}_m$. Assuming a harmonic oscillator model, the energy gap is approximated as $\Delta {\mathcal E}_m \approx m k_{\rm B} \theta_{\rm v}$, where $k_{\rm B}$ is the Boltzmann constant and $\theta_{\rm v}$ is the characteristic vibrational temperature. Although real molecules possess a finite dissociation limit, the summations over quantum numbers are extended to infinity to maintain consistency with the harmonic oscillator potential; this approximation is physically justified as the rapid decay of transition probabilities and population densities at high energy levels renders the contribution of terms exceeding the dissociation limit negligible. Then,
\begin{equation}
    Q_{\rm e-v}^{(m)} = N_{\rm e} \sum_{n=0}^{\infty} N_n k_{n \to n+m} \Delta {\mathcal E}_m
    \label{eq:cooling_def}
\end{equation}
where $N_{\rm e}$ is the electron number density, $N_n$ is the population density of the vibrational level $n$, and $k_{n \to n+m}$ is the rate coefficient for the electron-impact excitation from level $n$ to $n+m$. The corresponding macroscopic heating rate for the reverse superelastic processes is given by
\begin{equation}
    Q_{\rm v-e}^{(m)} = N_{\rm e} \sum_{n=0}^{\infty} N_{n+m} k_{n+m \to n} \Delta {\mathcal E}_m
    \label{eq:heating_def}
\end{equation}
where $k_{n+m \to n}$ denotes the rate coefficient for the de-excitation from level $n+m$ to $n$. To relate these forward and reverse rates, we apply the principle of detailed balance to each transition channel individually. The rate coefficient for the reverse process is related to the forward process by the exponential of the specific energy gap involved at the electron temperature $T_{\rm e}$:
\begin{equation}
    k_{n+m \to n} = k_{n \to n+m} \exp\left(\frac{m\theta_{\rm v}}{T_{\rm e}}\right)
    \label{eq:detailed_balance}
\end{equation}
Furthermore, we assume a Boltzmann distribution for the vibrationally excited states. For a system in vibrational equilibrium at the vibrational temperature $T_{\rm v}$, the population density $N_n$ of a state with energy ${\mathcal E}_n$ is proportional to its Boltzmann factor:
\begin{equation}
    N_n \propto \exp\left(-\frac{{\mathcal E}_n}{k_{\rm B} T_{\rm v}}\right)
    \label{eq:boltzmann_prop}
\end{equation}
Consequently, the ratio of populations between an upper state $n+m$ and a lower state $n$ is determined by the energy difference between them:
\begin{equation}
    \frac{N_{n+m}}{N_n} = \exp\left(-\frac{{\mathcal E}_{n+m} - {\mathcal E}_n}{k_{\rm B} T_{\rm v}}\right)
    \label{eq:pop_ratio}
\end{equation}
Consistent with the harmonic oscillator approximation utilized in the cooling definition, the energy levels are equally spaced by the vibrational quantum $k_{\rm B} \theta_{\rm v}$, such that the energy gap is
\begin{equation}
    {\mathcal E}_{n+m} - {\mathcal E}_n \approx m k_{\rm B} \theta_{\rm v}
    \label{eq:energy_gap}
\end{equation}
Substituting Eq.~(\ref{eq:energy_gap}) into Eq.~(\ref{eq:pop_ratio}) allows the Boltzmann constant $k_{\rm B}$ to cancel, relating the population of the upper state $N_{n+m}$ directly to the lower state $N_n$ via the characteristic vibrational temperature $\theta_{\rm v}$:
\begin{equation}
    N_{n+m} = N_n \exp\left(-\frac{m\theta_{\rm v}}{T_{\rm v}}\right)
    \label{eq:boltzmann}
\end{equation}
Substituting Eq.~(\ref{eq:detailed_balance}) and Eq.~(\ref{eq:boltzmann}) directly into the definition of the heating rate in Eq.~(\ref{eq:heating_def}) yields
\begin{equation}
    Q_{\rm v-e}^{(m)} = N_{\rm e} \sum_{n=0}^{\infty}  \left[
    N_n \exp\left(-\frac{m\theta_{\rm v}}{T_{\rm v}}\right)  k_{n \to n+m} \exp\left(\frac{m\theta_{\rm v}}{T_{\rm e}}\right)  \Delta {\mathcal E}_m
    \right]
    \label{eq:substitution}
\end{equation}
Rearranging the terms to group the temperature-dependent exponentials, we obtain
\begin{equation}
    Q_{\rm v-e}^{(m)} = N_{\rm e} \sum_{n=0}^{\infty}  \left[
    N_n   k_{n \to n+m} \exp\left(\frac{m\theta_{\rm v}}{T_{\rm e}}-\frac{m\theta_{\rm v}}{T_{\rm v}}\right)  \Delta {\mathcal E}_m
    \right]
    \label{eq:substitution_2}
\end{equation}
Since the exponential term in Eq.~(\ref{eq:substitution_2}) depends only on the channel order $m$ and the temperatures, it can be factored out of the summation, as follows:
\begin{equation}
    Q_{\rm v-e}^{(m)} =   \exp\left(\frac{m\theta_{\rm v}}{T_{\rm e}} - \frac{m\theta_{\rm v}}{T_{\rm v}}\right)   N_{\rm e} \sum_{n=0}^{\infty} N_n k_{n \to n+m} \Delta {\mathcal E}_m
    \label{eq:rearranged}
\end{equation}
The remaining terms (other than the exponential term) are identical to the definition of the cooling rate $Q_{\rm e-v}^{(m)}$ in Eq.~(\ref{eq:cooling_def}). Thus,
\begin{equation}
    Q_{\rm v-e}^{(m)} =       
    Q_{\rm e-v}^{(m)}
\exp\left(\frac{m\theta_{\rm v}}{T_{\rm e}} - \frac{m\theta_{\rm v}}{T_{\rm v}}\right)  
    \label{eq:rearranged_2}
\end{equation}
Consequently, the total vibrational-electron heating rate is expressed as the sum of these thermodynamically consistent components,
\begin{equation}
    Q_{\rm v-e} = \sum_{m=1}^\infty  Q_{\rm e-v}^{(m)} \exp\left(\frac{m\theta_{\rm v}}{T_{\rm e}} - \frac{m\theta_{\rm v}}{T_{\rm v}}\right)
    \label{eq:final_result}
\end{equation}
This formulation preserves the strict zero-net-transfer condition at thermal equilibrium ($T_{\rm e} = T_{\rm v}$) while accurately capturing the higher-energy kinetics of level-skipping transitions.

Central to this derivation are two key assumptions: the validity of the harmonic oscillator model and the existence of a Boltzmann distribution for vibrationally excited states. The latter is generally well-justified by the rapid rate of vibration-vibration (V-V) energy transfer, which typically thermalizes vibrational levels within approximately 50 collisions. In most hypersonic flow regimes, where the translational gas temperature is commensurate with or higher than the vibrational temperature, this fast V-V exchange robustly maintains a Boltzmann distribution. However, in specific non-equilibrium conditions where the gas temperature is significantly lower than the vibrational temperature---such as in the inter-pulse relaxation phase of plasma-assisted combustion (PAC) or laser-induced plasmas (LIP)---the distribution may instead relax to a Treanor distribution, accounting for the anharmonicity of the vibrational potential. While this represents a limitation, the Boltzmann assumption remains a standard and effective approximation for macroscopic fluid models, especially given the high computational cost of state-to-state kinetics.

\begin{figure}[!t]
    \centering
    \includegraphics[width=0.99\linewidth]{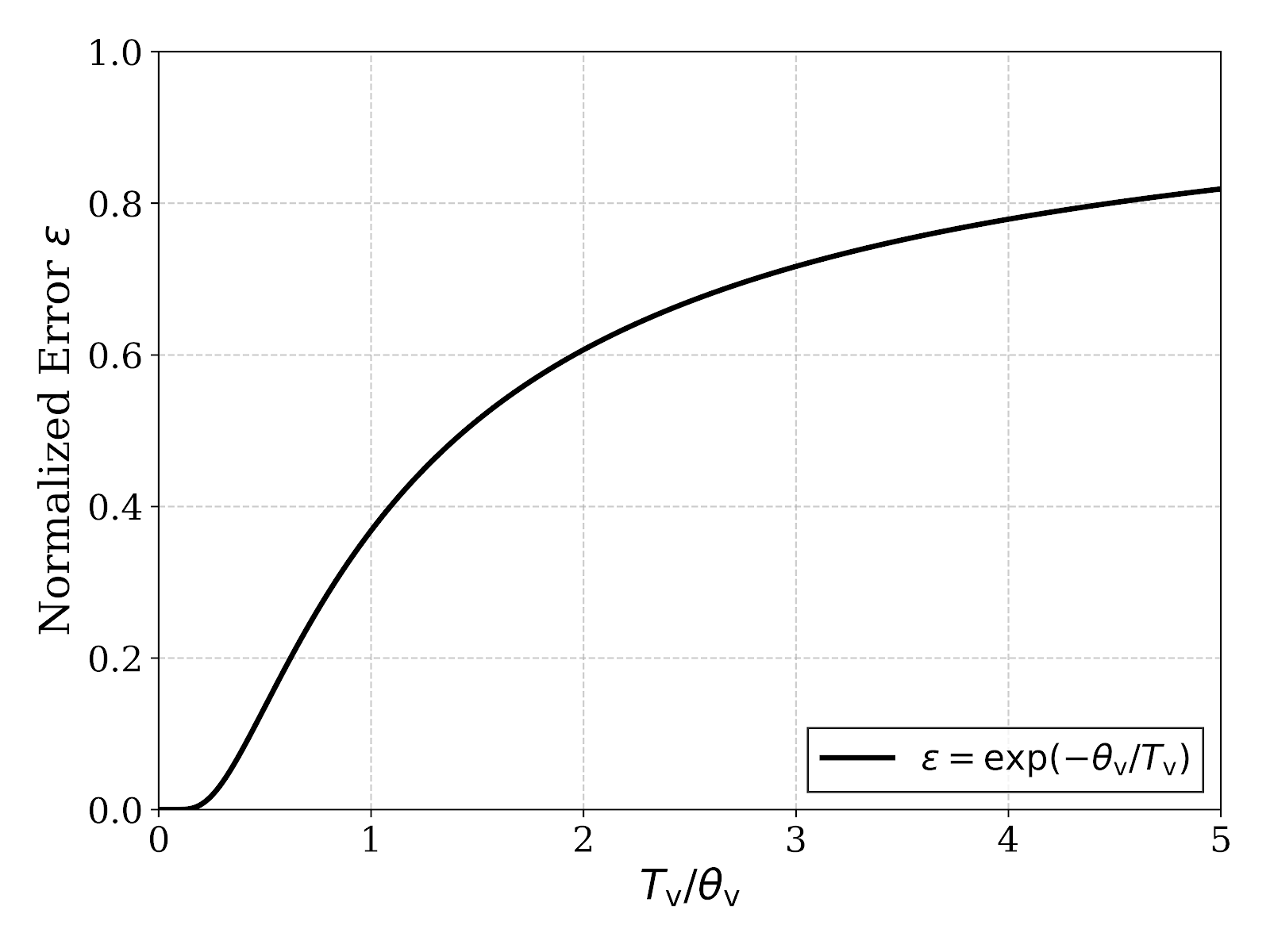}
    \figurecaption{Normalized error $\varepsilon$ in the Peters et al.\cite{jap:2019:peters} model  as a function of vibrational temperature. The error represents the fraction of electron heating flux neglected by omitting ``hot band'' transitions.}
    \label{fig:error_plot}
\end{figure}

The necessity of such a thermodynamically consistent closure becomes evident when contrasted with prior formulations, such as the model proposed by Peters et al.\cite{jap:2019:peters}. Their model's thermodynamic inconsistency is revealed by examining the ratio of electron heating to cooling at thermal equilibrium. As derived in their work, this ratio is:
\begin{equation}
    \left.\frac{Q_{\rm v-e}}{Q_{\rm e-v}}\right|_{\rm Peters} = \left(N_{\rm e}  {\sum_{n=1}^{\infty} N_n k_n^{\rm inv} \mathcal{E}_n}\right)\left(N_{\rm e}N {\sum_{n=1}^{\infty} k_n \mathcal{E}_n}\right)^{-1}
    \label{eq:peters_ratio}
\end{equation}
where $N_{\rm e}$ is the electron number density, $N$ is the total number density (including the ground state), ${\cal E}_n$ is the activation energy of the $n$th quantum level, $k_n$ is the forward excitation rate, and $k_n^{\rm inv}$ is the reverse rate derived from detailed balance. In the limit of thermal equilibrium ($T_{\rm e} = T_{\rm v}$), we apply the detailed balance relation $k_n^{\rm inv} = k_n \exp(n\theta_{\rm v}/T_{\rm e}) = k_n \exp(n\theta_{\rm v}/T_{\rm v})$. Substituting this into the numerator of Eq.~(\ref{eq:peters_ratio}), and expressing the number density of the $n$th level as $N_n = N_{n=0} \exp(-n \theta_{\rm v}/T_{\rm v})$, the exponential terms cancel exactly:
\begin{equation}
\sum_{n=1}^{\infty} N_n k_n^{\rm inv} \mathcal{E}_n
=
    \sum_{n=1}^{\infty} N_{n=0} \exp\left(-\frac{n\theta_v}{T_v}\right)  k_n \exp\left(\frac{n\theta_v}{T_v}\right) 
    \mathcal{E}_n = N_{n=0} \sum_{n=1}^{\infty} k_n \mathcal{E}_n
\end{equation}
Substituting this result back into Eq.~(\ref{eq:peters_ratio}), the summations cancel, and the heating-to-cooling ratio simplifies to the fraction of molecules in the ground state:
\begin{equation}
    \left.\frac{Q_{\rm v-e}}{Q_{\rm e-v}} \right|_{{\rm Peters},~T_{\rm e}=T_{\rm v}} =\frac{N_{n=0}}{N} 
    \label{eq:peters_ratio_2}
\end{equation}
For a Boltzmann distribution, this fraction corresponds to:
\begin{equation}
    \left.\frac{Q_{\rm v-e}}{Q_{\rm e-v}} \right|_{{\rm Peters},~T_{\rm e}=T_{\rm v}} = 1 - \exp\left(-\frac{\theta_{\rm v}}{T_{\rm v}}\right)
    \label{eq:peters_ratio_3}
\end{equation}
It is instructive to define a normalized error, $\varepsilon$, representing the deviation from thermodynamic consistency at equilibrium for the Peters et al.\ model:
\begin{equation}
    \varepsilon = 1 - \left.\frac{Q_{\rm v-e}}{Q_{\rm e-v}} \right|_{{\rm Peters},~T_{\rm e}=T_{\rm v}} = \exp\left(-\frac{\theta_{\rm v}}{T_{\rm v}}\right)
    \label{eq:error_def}
\end{equation}
Figure~\ref{fig:error_plot} plots this error as a function of the normalized vibrational temperature.
As illustrated in Fig.~\ref{fig:error_plot}, the error is substantial. Mathematically, it vanishes as $T_{\rm v} \to 0$ (where the population resides almost entirely in the ground state) but rises sharply with temperature. For conditions where $T_{\rm v} \gtrsim \theta_{\rm v}$, the model neglects more than 40\% of the requisite heating flux.

Physically, the form of Eq.~(\ref{eq:peters_ratio}) reveals that the missing energy corresponds exactly to the population fraction of vibrationally excited states ($n \ge 1$). This highlights the asymmetry in the formulation of the macroscopic rates: the cooling term (denominator) scales with the total number density $N$, effectively assuming all molecules contribute to energy loss. In contrast, the heating term (numerator) accounts only for superelastic transitions returning to the ground state ($n \to 0$), thereby ignoring ``hot band'' heating transitions between excited states (e.g., $n \to n-1$).

One might attempt to restore thermodynamic consistency in Eq.~(\ref{eq:peters_ratio}) by replacing the total number density $N$ in the denominator with the ground-state density $N_{n=0}$. While this modification would force the ratio to unity at equilibrium, it would physically correspond to neglecting electron cooling by vibrationally excited molecules. At elevated vibrational temperatures ($T_{\rm v} > \theta_{\rm v}$), these excited states become dominant contributors to the cooling flux; ignoring them would result in a significant underestimation of the total electron cooling rate.

Consequently, as $T_{\rm v}$ increases, the Peters et al.\ formulation artificially restricts the heat source to a shrinking fraction of the population proportional to $N_{n=0}$, while counting the entire population in the heat sink. This inconsistency is corrected in the present work by adopting the generalized thermodynamically consistent formulation. By enforcing the macroscopic detailed balance relation outlined in Eq.~(\ref{eq:final_result}), we implicitly account for the missing energy flux from these hot-band transitions, guaranteeing that $\varepsilon = 0$ (or $Q_{\rm v-e} = Q_{\rm e-v}$) exactly when $T_{\rm e} = T_{\rm v}$, while retaining the accuracy of the total cooling rates.

Ultimately, this generalized framework captures the critical heating pathways dictated by vibrationally excited states, sustaining the intense transition from high-energy discharge pulses to the soft plasma afterglow.

\section*{Acknowledgments}

The authors sincerely thank the anonymous reviewers for their constructive guidance. We are particularly grateful for the insightful suggestion to generalize the derivation for multi-quantum transitions, which resonated deeply and inspired the expanded scope of this work.

%\appendix
%\section{Appendix Test}
\footnotesize
\bibliography{all}
\bibliographystyle{aiaa2}
\end{document}